\documentclass[11pt]{amsart}
\usepackage{geometry}                
\usepackage{graphicx}
\usepackage{amssymb}
\usepackage{natbib}

 \newcommand{\bfE}{\mathbf{E}}

\newcommand{\bfJ}{\mathbf{J}}

\newcommand{\bfv}{\mathbf{v}}

\newcommand{\bfx}{\mathbf{x}}

\title[Implicit  discretization and  energy conservation for   Poisson-Boltzmann ]{Implicit temporal discretization and exact energy conservation for particle methods applied to the Poisson-Boltzmann equation}

\author[Giovanni Lapenta, Wei Jiang]{Giovanni Lapenta\\
Departement Wiskunde, University of Leuven, KU Leuven, Leuven, Celestijnenlaan 200B, 3001 Leuven, Belgium, European Union; giovanni.lapenta@kuleuven.be\\ Wei Jiang \\
School of Physics, Huazhong University of Science and Technology, Wuhan 430007, People's Republic
of China; physics.tame@gmail.com}

\begin{document}

\abstract We report on a new multiscale method approach for the study of systems with wide separation of short-range forces acting on short time scales and long-range forces acting on much slower scales. We consider  the case of the Poisson-Boltzmann equation that describes the long-range forces using the Boltzmann formula (i.e. we assume the medium to be in quasi local thermal equilibrium).
We developed a new approach where fields and particle information (mediated by the equations for their moments) are solved self-consistently. The new approach is implicit and numerically stable, providing exact energy conservation. We tested different implementations all leading to exact energy conservation. 
The new method  requires the solution of a large set of non-linear equations. We considered three solution strategies: Jacobian Free Newton Krylov, an alternative, called field hiding, based on hiding part of the residual calculation and replacing them with direct solutions and a Direct Newton Schwarz solver that considers simplified single particle-based Jacobian. The field hiding strategy proves to be the most efficient approach.}

\maketitle

\section{Introduction}
Matter in any form and state is characterized by the presence of particles interacting via fields.  At the most fundamental level, a quantum point of view is needed but as larger and larger portions of matter need to be studied, the attention shifts from the quantum level to the particle level (where matter is described as particles interacting via forces), to the macroscopic level (where matter is described as a continuum with properties defined everywhere in space). 

Perhaps the key effort in science and in particular in computational science is to design models able to describe or predict the properties and behavior of matter based on the knowledge of its constituents and their interaction: We consider here in particular the {\it particle approach} \citep{hockney-eastwood,frenkel2001understanding}. At the core of the method is the ability to use supercomputers to track millions or billions of particles to reproduce the behavior of molecules, proteins, genome and any type of matter. The fundamental complexity and challenge is that the evolution of such systems encompasses many scales \citep{de2008challenges}. 

Most first-principle methods are required to resolve the smallest scales present not because the processes there are important but because otherwise the method would fail due to numerical instabilities \citep{hockney-eastwood}. Particle models must account for the presence of long and short-range interaction that acts on different temporal scales. The contribution from short-range forces can be computed for each particle considering only the others within a short distance. But the long-range forces require a global approach for the whole system. Different methods have been designed \citep{frenkel2001understanding}. One class of approaches, called the particle-particle particle-mesh (P3M)  \citep{eastwood1984p3m3dp} and the more recent Particle-mesh Ewald (PME) summation \citep{essmann1995smooth}, introduce a mesh for the long-range interaction. 

In practical terms the short-range calculation is less demanding as it involves only the particles within a cell selected to meet the short-range forces, but the long-range interaction is a costly mesh operation involving the whole domain and requiring the solution of an elliptic equation of the Poisson-type. 

Additionally, short-range calculations can be effectively and easily implemented on modern massively parallel computers and especially on GPU-based computers. 
Long-range calculations require communication among processors as distant areas of the system still need to exchange information. 

For these reasons, new methods have been developed to take advantage of a fundamental difference between short and long range interaction: by virtue of their long range, forces acting over long distances affect the system on slower scales. The reversible reference system propagator algorithms (RESPA) \citep{tuckerman1992reversible} use more frequent updates for the short-range forces while the long-range forces can be recomputed only intermittently. These approaches are based on decoupling the force calculation for the different contributions to the interaction (short and long range) from the particle mover: the mover calls the force calculation at different time intervals depending on the time scale typical of the force. Long-range interactions are  computed less frequently reducing the cost. 

But it is possible to go even further and reduce drastically the frequency of long-range force calculation if the particle mover and the force field computation are not decoupled: using implicit   methods \citep{brackbill2014multiple}. This is the approach we propose here. Our new approach allows us to model the coupled particle-motion equations and long-range Poisson equation within a self-consistent set of mesh equations that can be solved with much longer time steps. The key innovation of the implicit approach is, however, its capability to conserve energy exactly and to machine precision, a property that remains elusive for explicit methods~\citep{salmon1994skeletons,hockney-eastwood,birdsall-langdon,GrigoryevPIC}.

The downside of implicit method is of course the need for iterative solvers. We consider here the case of solvation \citep{im1998continuum} where the system is described by the of Poisson-Boltzmann system \citep{fogolari2002poisson}, a highly non-linear system of coupled equations. We have to rely on the Newton method for its solution, but we explore two practices to reduce the cost: a method that removes part of the operations (solution of the field equations) from the main non-linear iterations (we refer to this approach as \textit{field hiding}) and one where the Jacobian of the Newton iteration is drastically simplified (we refer to this approach as \textit{Newton-Krylov-Schwarz}). 

The method presented is  tested in a standard problem where complex physics developed. Energy is observed to remain conserved within arbitrary precision. The different solution strategies all converge to the correct solution but the field hiding approach is observed to be the most cost effective.

The paper is organized as follows. Section 2 describes the model problem we use: implicit solvation, while Section 3 describes its mathematical implementation in the Poisson-Boltzmann system. Section 4 describes two possible temporal discretization: the standard explicit method widely used in the literature and our new implicit approach. Section 5 describes a key novelty of our approach, it ability to conserve energy exactly to machine precision. Section 6 describes he solution strategy for the implicit approach, based on the Newton-Krylov method and on two alternative approaches that can riduce the computational cost. Section 7 present results for a specific case, showing exact energy conservation. Section 8, instead, focuses on the performance of the different solution strategies. Finally Section 9 outlines the conclusions of our work.

\section{Model Problem: Implicit Solvation}
Our goal is the application of  implicit  methods to improve the handling of long-range forces in multiple scales particle simulations. We consider as target problem a system of interacting atoms or ions immersed in a matrix.  

One example guided the progress. Atoms, molecules and particles immersed in media, like water, or in other matrices are present in many materials and in particular in all biomaterials. Treating media as a collective entity is a more effective approach than trying to model all atoms in the medium. We based our approach on  the Poisson-Boltzmann (PB) equation that treats the electrostatic behavior of the media as a continuum model introducing an electrostatic field solved on a grid \citep{im1998continuum, lu2008recent}. 

In these cases, our approach is to treat the motion of the particles of the solute using a mesh to handle long-range forces:  the use of a Poisson-Boltzmann model for long range interactions and the effects of solvation. The particles in the system evolve under action of short-range forces and the long-range effect of a potential field computed from the Poisson equation. 

Our approach is different from the  existing state of the art based on multiscale RESPA \citep{tuckerman1992reversible} particle integrators.  The implicit  method changes the computational cycle by replacing a Verlet-style explicit alternation of particle motion and force computation with a coupled system of equations for both particles and fields. 

In a standard particle-based method, the particles are moved for a short distance in the forces computed at the start of the time step. Then the forces are recomputed and the particles are moved again. This requires the use of very small time steps that resolve the smallest scale present to avoid numerical instability \citep{birdsall-langdon,frenkel2001understanding}. The implicit method, instead, solves the coupled non-linear system for particles and fields self-consistently. This allows one to move particles for longer distances in a single time step, thereby allowing the study of much larger systems for much longer times \citep{Brackbill:1985}. 

The approach is ideally suited for situations where the long-term evolution is of interest and the short scales can be averaged over. For example, in the case of macromolecules, this would allow to quickly allow structural changes. 

 Recent work has demonstrated mathematically and in practice that the approach conserves energy exactly, to round off \citep{markidis2011energy,lapenta2011particle,chen2011energy}. This is a sore point in particle simulations. Multiscale methods like the widely used RESPA suffer from numerical limitations that can lead to unphysical building up of energy in the system, thereby giving rise to drifts in average properties and inaccurate sampling \citep{morrone2010molecular}. This effect is completely eliminated by the implicit energy conserving approach proposed here.

\section{The Poisson-Boltzmann model}
In our study we use the Poisson-Boltzmann model as basis of our investigations. We briefly review the technique we use to provide the background on the new methods we developed

The Poisson Boltzmann equation describes the electrochemical potential of ions immersed in a medium (e.g. water). 
\begin{equation}
\epsilon_0\nabla^2 \varphi = -\rho_i +\rho_m 
\end{equation}

The charge on the right hand side can come from two different sources: ions immersed in the fluid $\rho_i$ and the charges present in the medium itself that responds to the evolving  conditions $\rho_m$. A classic example is the dispersion in water that responds by polarizing its molecules.

The ions are described using computational particles. Usually these particles are not infinitesimal in size but rather are assumed to be of finite size, with a prescribed shape. 

The medium charge, conversely, is defined by a continuum model. In the case of the Boltzmann model, the response of the medium is based on the Boltzmann factor: 
\begin{equation}
\rho_m = \rho_{0s} \exp(-q_m\varphi/k_BT_m)
\end{equation}
where $k_B$ is the Boltzmann constant, defined by a temperature $T_m$ and an asymptotic density at regions where the potential vanishes is  $\rho_0$.

We can consider a more general non-linear response of the medium and generalize the Poisson-Boltzmann model to:
\begin{equation}\label{eq:pboltz}
\epsilon_0\nabla^2 \varphi = -\rho_i +f( \varphi) 
\end{equation}
for a generic non-linear function $f$. For example, a often-used formulation leads to: $f(\varphi)=\sinh(\gamma \varphi)$ where $\gamma$ is a normalization parameter \citep{honig1995classical,koehl2006electrostatics}.

The equation above forms our model for the medium. For ions, the equations of motion for the position $\bfx_p$ and velocity $\bfv_p$ of each ion are simply given by Newton's equations:
\begin{equation}
\begin{array}{l}
\displaystyle  \frac{d\bfx_p}{dt}=\bfv_p\\ \\
\displaystyle  \frac{d\bfv_p}{dt}=\frac{q_i}{m_i}\bfE(\bfx_p)
\end{array}
\end{equation}
where the mass and charge of the ionic species are indicated as usual. 

The electric field acting on the ions comes from two sources: the direct interaction among the ions and the interaction mediated by the medium. This is in the spirit of the P$^3$M method that splits the short-range and long-range interaction \citep{eastwood1984p3m3dp}. 
The direct interaction is coming from particle-particle forces:
\begin{equation}
\displaystyle \bfE(\bfx_p)=-\sum_{p^\prime}\nabla_{\bfx_{p^\prime}} V(\bfx_p-\bfx_{p^\prime})
\end{equation}
where $V$ is the inter-particle potential and the summation is to all particles (avoiding self-forces).

The interaction mediated by the medium is obtained solving the Poisson equation on a grid of points $\varphi_g=\varphi(\bfx_g)$. From the grid potential, we can compute the grid electric field and project it to the particles:
\begin{equation}
\displaystyle \bfE(\bfx_p)=\sum_g\bfE_g W(\bfx_p-\bfx_g)
\end{equation}
where the summation is over all points of the grid. For typical interpolation functions, $W(\bfx_p-\bfx_g) $ only few cells contribute to the particles. We use here for interpolation b-splines of order 1 \citep{de1978practical}.  

The peculiarity of particle  methods is the use of  interpolation functions $W \left( {x}_g - \bar{x}_p \right)$ ($g$ the generic grid point, center or vertex) to describe the coupling between particles and fields. 
We use here  the Cloud-in-Cell approach, the interpolation function reads~ \citep{hockney-eastwood}:
\begin{equation} \label{eq:interp}
W\left( {x}_g - \bar{x}_p \right)=\begin{cases}
1 - \frac{| {x}_g - \bar{x}_p |}{\Delta x}, & \text{if } | {x}_g - \bar{x}_p | < \Delta x \\
0, & \text{otherwise. } \\
\end{cases}
\end{equation}

The equations of particle motion and the Poisson equation for the field are a non-linear coupled system of equations. The goal here is to understand how to deal with this coupling so that energy is conserved.

\section{Explicit and Implicit Discretization of the Poisson-Boltzmann Equation}
The set of equations above can be discretized in time explicitly or implicitly. The explicit approach solves in sequence  in each time interval $\Delta t$  the two equations: first it solves one assuming the other frozen and then vice-versa. 

Considering the equations for the computational particles with charge  $q_{p}$ and mass $m_{p}$, the explicit approach uses the so-called leap-frog (or Verlet) algorithm: 
\begin{equation} \label{eq:PICeq}
\begin{split}
&{\bfx}_p^{n+1} = {\bfx}_p^n + {\bfv}_p^{n+1/2} \Delta t \\
&{\bfv}_p^{n+1/2} = {\bfv}_p^{n-1/2} + \frac{q_p}{ m_p} \Delta t \bfE^{n}(\bfx_p^n).   
\end{split}
\end{equation}
The new particle position can be directly computed from the old electric field. After moving the particles, the new electric field can then be computed form Poisson's equation.

On the contrary, the implicit method uses a formulation where field and particles are advanced together within an iterative procedure where at each time step the field equation and the particle equations are solved together.
The implicit mover used here is: 
\begin{equation} \label{eq:PICeqimpl}
\begin{split}
&{\bfx}_p^{n+1} = {\bfx}_p^n + \bar{{\bfv}}_p \Delta t \\
&\bar{{\bfv}}_p = {\bfv}_p^n + \frac{q_p}{2 m_p} \Delta t \bar{{\bfE}}_p.   
\end{split}
\end{equation}
where quantities under bar are averaged between the time step $n$ and $n+1$ (e.g. $\bar{{\bfv}}_p = ({\bfv}_{p}^{n}  +{\bfv}_{p}^{n+1})/2$). The new velocity at the advanced time is then simply:
\begin{equation} \label{eq:ECPICmover_av}
{\bfv}_p^{n+1} = 2 \bar{{\bfv}}_p - {\bfv}_p^n.
\end{equation}   
The electric field  $\bar{E}_p$ is the  electric field acting of the computational particle  at the mid-time $\bar{{E}}_p = ({E}_{p}^{n}  +{E}_{p}^{n+1})/2$.

The electric field is computed from the direct interaction among particles and from the Poisson equation. Focusing on the latter,  to reach an exactly energy conserving scheme, we reformulate the equation using explicitly the equation of charge conservation:
\begin{equation} \label{eq:chcons}
\frac{\partial \rho_i}{\partial t} = - \nabla \cdot  \bfJ_i
\end{equation}
We take the temporal partial derivative of the Poisson's equation (\ref{eq:pboltz}) and substitute eq. (\ref{eq:chcons})
\begin{equation} \label{eq:field}
-\epsilon_{0}  \nabla \cdot \frac{\partial \bfE}{\partial t} =   \nabla \cdot \bfJ_i +\frac{\partial f}{\partial \varphi} \frac{\partial \varphi}{\partial t}
\end{equation}
The last term can then be interpreted as the divergence of the current of charge in the medium:
\begin{equation} \label{eq:field2}
\nabla \cdot \bfJ_m = -\frac{\partial f}{\partial \varphi} \frac{\partial \varphi}{\partial t}
\end{equation}

The result above allows us to rewrite the Poisson-Boltzmann model as:
\begin{equation} \label{eq:field3}
 \nabla \cdot \left( \epsilon_0 \frac{\partial \bfE}{\partial t} -   \bfJ_i +\bfJ_m \right) =0
\end{equation}
that can be solved by numerical discretization:
\begin{equation} \label{eq:field4}
{\mathcal{D}} \cdot \left( \epsilon_0 \Delta \bfE_g - \Delta t   (\bar{\bfJ}_{ig} +\bar{\bfJ}_m )\right) =0
\end{equation}
where $\mathcal{D}$ is the discretized divergence operator and $\Delta \bfE_g$ is the variation of $\bfE_g$ during the time step. This equation needs to be solved directly in this form. Formally, the divergence operator can be inverted (provided suitable boundary conditions) leading to:
 \begin{equation} \label{eq:field5}
\epsilon_0 \Delta \bfE_g = \Delta t  ( \bar{\bfJ}_{ig} -\bar{\bfJ}_m) 
 \end{equation}
This equation is our basis for proving the exact conservation but it is not of direct practical use. 

To solve the equation above one needs first to invert eq.(\ref{eq:field2}), a task of comparable complexity to inverting eq. (\ref{eq:field4}). We prefer the latter because it allows more easily to set boundary conditions. The solution for the field equations is then found  assuming the electric field can be expressed via the potential using a discretized  gradient $\mathcal{G}$: 
\begin{equation} \label{eq:newE}
\Delta \bfE_g \equiv \bfE^{n+1}_g - \bfE^{n+1}_g = -\mathcal{G} \Delta\varphi_g^n
\end{equation}
and inverting the system of eqs. (\ref{eq:field4}).

The time dependent Poisson formulation is Galilean invariant and does not suffer from the presence of any curl component in the current \citep{chen2011energy}. In electrostatic systems, the current cannot develop a curl because such curl would develop a corresponding curl of the electric field, and in consequence electromagnetic effects. The formulation above prevents that occurrence since it is based on the divergence of the current and any  curl component of J is eliminated. This is not an issue in 1D but it is central in higher dimensions.

\section{Energy Conserving Fully Implicit Method (ECFIM)}

There are several energy channels and they need to balance exactly for energy to be conserved. Let us begin by the particles. Their energy change can be computed easily multiplying the momentum equation by the average between new and old velocity over the time step:
For the explicit scheme this leads to:
\begin{equation}
\sum_p \frac{m_p}{2}\left( ({\bfv}_p^{n+1/2})^2 - ({\bfv}_p^{n-1/2})^2 \right) =\sum_g\sum_p \frac{q_p}{2} \Delta t \bfE^{n}(\bfx_g^n) W(\bfx_p-\bfx_g)\cdot ({\bfv}_p^{n+1/2} + {\bfv}_p^{n-1/2}) 
\end{equation}
where we can recognize the current as 
\begin{equation}
\bfJ_{ig} V_g = \sum_p q_p  W(\bfx_p-\bfx_g) ({\bfv}_p^{n+1/2} + {\bfv}_p^{n-1/2}) 
\end{equation}
where $V_g$ is the volume of the cell. 
The particle energy balance then becomes: 
\begin{equation}\label{eq:balpar}
\sum_p \frac{m_p}{2}\left(({\bfv}_p^{n+1/2})^2 - ({\bfv}_p^{n-1/2})^2 \right) =\Delta t \sum_g\bfE_g^n \cdot \bfJ_{ig}  V_g
\end{equation}

Similarly, for the implicit method, the energy balance is:
 \begin{equation}
 \sum_p \frac{m_p}{2}\left(({\bfv}_p^{n+1})^2 - ({\bfv}_p^{n})^2 \right) =\sum_g\bar{\bfE}_g W(\bfx_p-\bfx_g)\cdot \bar{\bfJ}_{ig} V_g
 \end{equation}
where the current is computed as:
 \begin{equation}
\bar{ \bfJ}_{ig} V_g = \sum_p q_p  W(\bfx_p-\bfx_g) \bar{\bfv}_p 
 \end{equation}
The particle energy balance then becomes: 
\begin{equation}\label{eq:balparimp}
\sum_p \frac{m_p}{2}\left(({\bfv}_p^{n+1})^2 - ({\bfv}_p^{n})^2 \right) =\Delta t \sum_g\bar{\bfE}_g \cdot \bar{\bfJ}_{ig}  V_g
\end{equation}

There is a crucial difference between implicit and explicit methods. The change in particle energy in the explicit method is given by the product of the old electric field with the current based on the new particle velocity.  That electric field was computed with the current from the old time step. There is then an inconsistency between the current used to advance the field and that coming out of the motion of the particles. there is a temporal delay of one time step.  The result is that energy is not conserved. In the implicit method, instead, the electric field and the current are computed at the same time level in both particle equations and field equation. Energy is conserved exactly. 

To prove the last point, let us now multiply eq. (\ref{eq:field5}) by $\bar{\bfE}_g$:
  \begin{equation} \label{eq:balfieldimp}
\sum_g \frac{\epsilon_0V_g}{2} \left ( (E_g^{n+1})^2 -(E_g^{n})^2 \right )= \Delta t   \sum_g V_g \bar{\bfJ}_{ig} \cdot \bar{\bfE}_g - \Delta t \sum_g V_g \bar{\bfJ}_m \cdot \bar{\bfE}_g
 \end{equation}

We recognize in this balance the variation of electric field energy on the left. The right hand side has the energy exchange between particles and fields and between medium and fields. 

The requirement of energy conservation is that the energy exchange between particles and fields computed form the particles, eq. (\ref{eq:balparimp}) exactly equals the exchange between particles and fields computed form the fields, eq. (\ref{eq:balfieldimp}). Inspection immediately shows this to be the case and energy is indeed exactly conserved for the implicit scheme.

Note that this conclusion holds with respect to both particle-based and mesh-based force computation. Above we focused on the particle-mesh component. But energy conservation of the particle-particle follows directly form the formulation in terms of inter-particle potential, provided the derivatives of the potential are taken also implicitly. To this end we discretize the gradient operator as follows:
\begin{equation}
 E_\alpha(\bfx_p)=-\frac{1}{\Delta t}\frac{ V(\bfx_p^{n+1}-\bfx_{p^\prime}^{n+1}) -V(\bfx_p^n-\bfx_{p^\prime}^n)}{\bar{v}_{\alpha p} -\bar{v}_{\alpha p^\prime} }
\end{equation}
where $\alpha$ is the direction in $\Re^3$.
The anti-symmetry of the expression above ensures energy and momentum conservation for the PP part of the computation \citep{GrigoryevPIC}.

\section{Newton-Krylov Solvers}
The implicit method described above produces a set of non linear equations, formed by the momentum equation of each particle (eq.(\ref{eq:PICeqimpl})) and the Poisson-Boltzmann equation for the variation of the potential (eq.(\ref{eq:field4})):
\begin{equation}
\left\{
\begin{array}{l}
\displaystyle\bar{{\bfv}}_p - {\bfv}_p^n - \frac{q_p}{2 m_p} \Delta t \bar{{\bfE}}_p =0
 \\ \\
{\mathcal{D}} \cdot \left( \epsilon_0 {\mathcal{G}}\Delta \varphi_g + \Delta t   (\bar{\bfJ}_{ig} +\bar{\bfJ}_m )\right) =0
\end{array}\right.
\label{eq:summary}
\end{equation}
These two residual equations are supplemented by a number of definitions: the calculation of the current from the particles and the calculation of the particle's electric field form the potential computed on the grid. These steps are part of the residual calculation but are not unknowns proper. The unknowns are $\Delta \varphi_g$ and  $\bfv_p$. All other quantities can be considered as derived constitutive relations that are not part of the unknowns variable of the coupled non linear system. The position can be computed immediately and linearly once the velocity is known. Once the position is known the current and density can be computed directly again linearly. As a consequence the only two sets of equations for the coupled non linear system are that for  $\varphi_g$ and  $\bfv_p$.

To solve this non linear system we use the \textit{Jacobian-Free Newton Krylov (JFNK)} approach\citep{kelley,knoll2004jacobian}. In this approach, the system of non linear equations is solved with the Newton method. Each iteration of the Newton method requires us to solve the linearized problem around the previous iteration for the solution. In  JFNK, this step is completed numerically using a Krylov solver (we use GMRES) where the Jacobian is not directly computed but rather only its products with Krylov vectors are computed. In practice, this means that the Jacobian never needs to be formulated or completely computed and only successive realizations of the residual need to be available.

The advantage is that the JFNK method can be used as a black box that takes as input a method that defines the residual  and as output it provides the solution. Many effective JFNK packages are available in the literature and most computing environment provide one. We use matlab.

\begin{figure}[h]
	\noindent\includegraphics[width=.45\columnwidth]{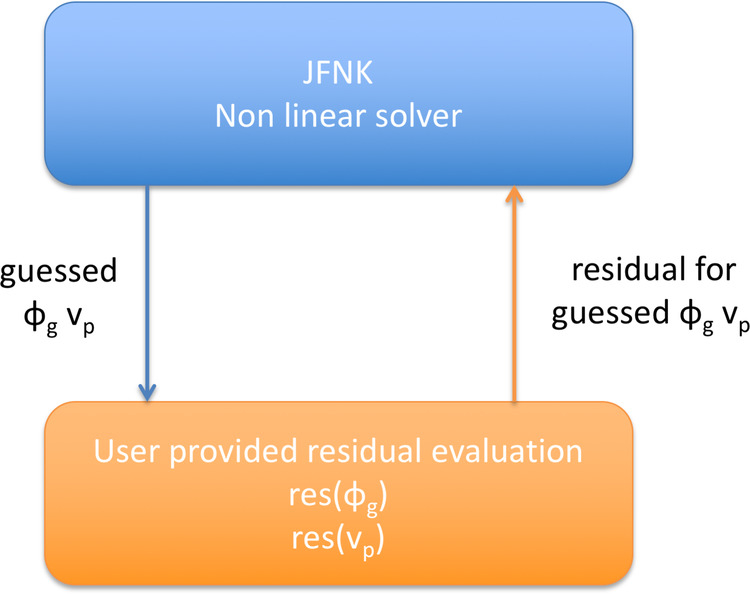}
	\caption{JFNK approach: the method interacts with the user provided residual evaluation that must supply to JFNK the residual error in the equations for a given guessed unknown variable. At convergence the residual will be smaller than a prescribed tolerance.}
	\label{nolinear1}
\end{figure}

In our case then the two residual equations are that for $\varphi_g$ and  $\bfv_p$. JFNK provides a sequence of guesses  for $\varphi_g$ and  $\bfv_p$ produced by the Newton iteration strategy, the user needs to provide a method for the residual evaluation given the guess. The final output is the  converged solution for $\varphi_g$ and  $\bfv_p$ that makes the residual smaller than a prescribed tolerance. The size of the problem treated by JFNK is  equal to the number of particle unknowns ($3N_p$ one for each velocity component) and of the potential unknowns ($N_g$). The approach is indicated in fig.~\ref{nolinear1}

The JFNK uses  two types of iterations, the inner Krylov iteration for the Jacobian equation and the outer Newton iteration. What counts at the end is how many residual evaluations are required for convergence. The larger the number of evaluations, of course, the larger the computational effort. 

Besides the direct implementation  described above, in our recent research we have explored other alternatives~\cite{siddicomparison}, possibly preferable in the strategy to reduce the memory requirements or the computational time.   These methods modify the residual equations, pre-computing part of the solution to eliminate part of the complexity of the non linear coupled system so that the JFNK method can converge more easily.

The first approach proposed in the context of implicit particle methods is that of \textit{particle hiding (PH)} \citep{kim}. In particle  hiding,  the unknowns of the problem are only the values of the potential on the grid points at the new time level and the JFNK solver computes only the residual of the field equations. The particle equations of motion are calculated by a separate Newton-Raphson method and embedded in the field solver as function evaluations.  When the JFNK provides a guess of the new potentials, particle positions and velocities are computed consistently with the electromagnetic field guessed, their current are then passed back to the residual evaluation for computing the field residuals required by JFNK. 

\begin{figure}[h]
	\noindent\includegraphics[width=.45\columnwidth]{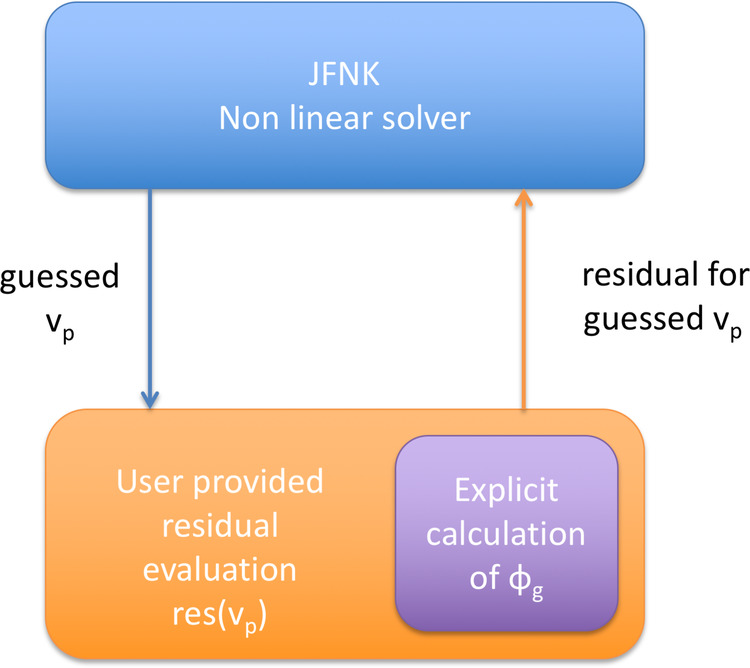}
	\caption{JFNK approach with field-hiding (FH): the user now computes part of the solution (hiding part of the calculation from the JFNK solver) reducing the burden carried by the JFNK solver itself. Faster performance can be achieved if properly implemented. With HH, the user solves the explicit field equation outside the Newton-Krylov iteration. }
	\label{nolinear}
\end{figure}

In essence the idea is that if the JFNK provides a guess for the electric field, the solution of the particle equations of motion does not require any non linear iterations: the particles can be moved in the guessed fields, no iteration needed. In this sense the particles become a constitutive function evaluation. Each time the JFNK requires a residual evaluation for  a guessed potential (and consequently electric field) in each grid point, the particles are first moved, the density and current are interpolated to the grid and the residual equations for the electric field can then be computed. This approach reduces dramatically the number of non linear equations that are solved (only $N_g$ potential equations), but at the cost of moving the particles many times, once for each residual evaluation. The main advantage of this method is the reduction of the memory requirements of the Krylov method because the particles have been brought out of the Krylov loop and only the filed quantities matter when computing the memory requirement.  This approach is especially suitable to hybrid architectures \citep{chen2012efficient} and can be made most competitive when fluid-based preconditioning is used \cite{chen2014fluid}. However, in the present case of a low dimensionality problem run on standard CPU computers, particle hiding is neither needed nor competitive.

In a recent study, we proposed an alternative approach:  \textit{field hiding (FH)} \cite{siddicomparison}. As the name suggests, the crucial difference is that in FH the JFNK operates directly on the particle mover, making now the field computation part of the residual evaluation. When the Newton iteration produces a guess of the particle velocities, the fields can be immediately computed via an evaluation of the current from the particles and solving the field equations. Given that typically there are at least two orders of magnitude fewer fields than particle unknowns, the cost of a field evaluation is typically very small compared with moving particles. 

The advantage of FH is that the JFNK method operates directly on the most sensitive part of the system, the particles. When FH and PH are compared, the most striking difference is the much lower number of Newton iterations needed for FH \cite{siddicomparison} (it should be pointed out that fluid preconditioning can change this result \cite{chen2014fluid}).  The reason for this result is that in PH the JFNK is trying to converge using a much less sensitive leverage. By acting directly on the particles, FH lets JFNK operate on the levers that are most sensitive. The fields at one particle depend only on the fields of the nearby nodes, while the fields in the system depend on all the particles and their motion.  In other words, the field equation is elliptic, coupling the whole system. As a consequence, converging a Newton method on the fields  requires many more iterations, and in each one the particles need to be moved. For this reason the present study focused on FH.

In Ref.~\cite{siddicomparison} a third option is also proposed: replacing the JFNK method for the particle residual in the FH strategy with a {\it Direct Newton-Schwarz (DNS)} approach. In this case, the equation for each particle is iterated independently of the others, assuming a Schwarz-like decomposition of the Jacobian. The idea is that particle-particle coupling is of secondary importance to particle-field coupling in terms of convergence of the scheme. In the field-hiding approach the full Jacobian of the residual of particle $p$ depends not only on the particle $p$ itself but also on the others:
\begin{equation}
J^{FH}_{pp^\prime}=\frac{\partial R_p}{\partial v_{p^\prime}}
\end{equation}  
The dependence of the residual of particle $p$, $R_p$ on the velocity of other particles is mediated by the fields that are hidden in the field solver. They are hidden but are non-zero. In the Direct Newton-Schwarz approach proposed by Ref. \cite{siddicomparison}, this coupling is approximated with a Picard iteration and the Jacobian of each particle is approximated as:
\begin{equation}
J^{DNS}_{pp^\prime}=\delta_{pp^\prime}\frac{\partial R_p}{\partial v_{p}}
\end{equation}
The formulas above are for the simpler 1D case for simplicity of notation but they are valid in 3D by interpreting $v$ as all components of the velocity. In 3D, the Jacobian of each particle is just a 3x3 matrix and there is no need for Krylov solvers to invert it: it can be inverted directly. For this reason we call the method Direct Newton Schwarz (rather than Newton Krylov Schwarz). 

The DNS is based on a strong simplification of the Jacobian and it is subject to the risk that the assumption might become invalid and the Picard iteration might be slow or even stall. We will see that indeed this is the case as the simulation size increases.  

Regardless of the implementation of the JFNK, all the methods considered conserve energy exactly, to machine precision (provided the Newton method convergence is sufficiently tight) and are absolutely identical in the accuracy of the results produced for the cases tested. They only differ in the computing performance. We focus then on comparing the different implementations in terms of CPU time used.

\section{Results} 

To test energy conservation and the computational implementation of the JFNK methods, we report the result of one sample problem. We initialize a simulation with two streams of positive ions immersed in a uniform solvating medium of density $\rho_0$, behaving according to the  Boltzmann factor (for negative electrons of charge $-e$):
\begin{equation}
	f(\varphi)= \exp(e\varphi/kT)
\end{equation}
with uniform medium temperature T (in general different from the kinetic temperature of the ions). In this case, the non linear Botlzmann term gives a current:
\begin{equation} 
\nabla \cdot \bfJ_m = -\frac{e}{kT}\exp(e\varphi/kT) \frac{\partial \varphi}{\partial t}
\end{equation}
Two possible options can be implemented. In the first (Method 1), the non-linear term is computed at the middle of the time step, to retain second order accuracy:
\begin{equation} 
\mathcal{D} \cdot \bar{\bfJ}_m = -\frac{e}{kT}\exp(e\bar{\varphi_g}/kT) \frac{\varphi_g^{n+1}-\varphi_g^{n}}{\Delta t}
\end{equation}
This approach leads to a non-linear field equation because $\bar{\varphi_g} = (\varphi_g^{n+1}-\varphi_g^{n})/2$. But this poses no particular complexity since the overall system made by particles and fields is non-linear anyway and this non-linearity is handled by the JFNK. Of course, the extra non-linearity might result in more iterations, and we will see in the results that this is indeed the case.

In the second (Method 2), the non-linearity is simplified with a time decentering at the beginning of the time step:
 \begin{equation} 
\mathcal{D} \cdot \bar{\bfJ}_m = -\frac{e}{kT}\exp(e\varphi_g^n/kT)\frac{\varphi_g^{n+1}-\varphi_g^{n}}{\Delta t}
 \end{equation}
 avoiding non-linearity in the field equation for $\varphi_g^{n+1}$.
 
Both choices conserve energy exactly because they just express differently the energy exchanged with the medium but do not affect the exchange of energy between particles and fields.  Method 2 is simpler to compute and as it will be shown below it still retains a comparable accuracy in practice. For that reason it is preferable in the case considered. This conclusion is valid for the present case and it might not hold in other problems. The first approach insuring non-linear consistency to second order might be advantageous in problems where non-linear balance between opposing terms are present \citep{knoll2004jacobian}: this is not the case here.

We present the results using adimensional quantities. The 1D  system has size  $L=20\pi$ run for a total time of $\omega_p T= 250$ using a total number of $800$ cycles. The system is discretized in $256$ cells using $40000$ particles.  This test is just a proof of principle and it is not meant as a production run for a full code.

The two beams have speed $v_b=\pm 0.2$ with a thermal spread of $V_{th}=0.01$. The evolution leads to the formation of electrostatic shocks (sometimes referred to as double layers) \citep{forslund1970formation}. This is a sharp difference from the case of two-stream instabilities of electrons. Here the physics is completely different because the medium described by the Boltzmann factor has a strong impact on the evolution. 

For the Boltzmann factor we choose the temperature corresponding to a Debye length of $\lambda_{D}=\sqrt{\epsilon_0 kT/ne^2}=10^{-3} L$ (i.e. $e/kT =253.3030$). Note that physically this is the electron temperature, in principle completely unrelated to the ion kinetic temperature that determines the thermal speed of the ions.

\begin{figure}[h]
	\centering
	\includegraphics[width=\textwidth]{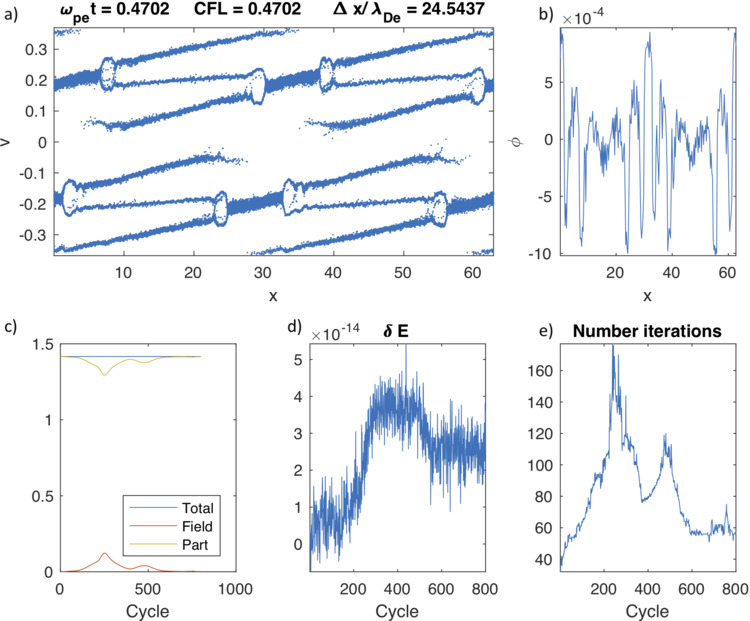}
	\caption{Method 1: Summary of a benchmark calculation, the final state is shown. Panel a: phase space (particle positions and velocities). Panel b: the potential $\varphi$. Panel c: Evolution of the energy. Panel d: Error in the conservation of energy. Panel e: Number of iterations needed for the NK iteration.}
	\label{fig:vanillajfnk}
\end{figure}

Figure \ref{fig:vanillajfnk}-a shows the final state of the run for Method 1. As can be observed, the evolution leads to the formation of double layers (sometimes referred to as electrostatic shocks)~\citep{forslund1970formation}. Figure \ref{fig:vanillajfnk}-b shows the  electrostatic potential at the end of the run. Sharp potential jumps are present in correspondence with each shock.  

\begin{figure}[h]
	\centering
	\includegraphics[width=\textwidth]{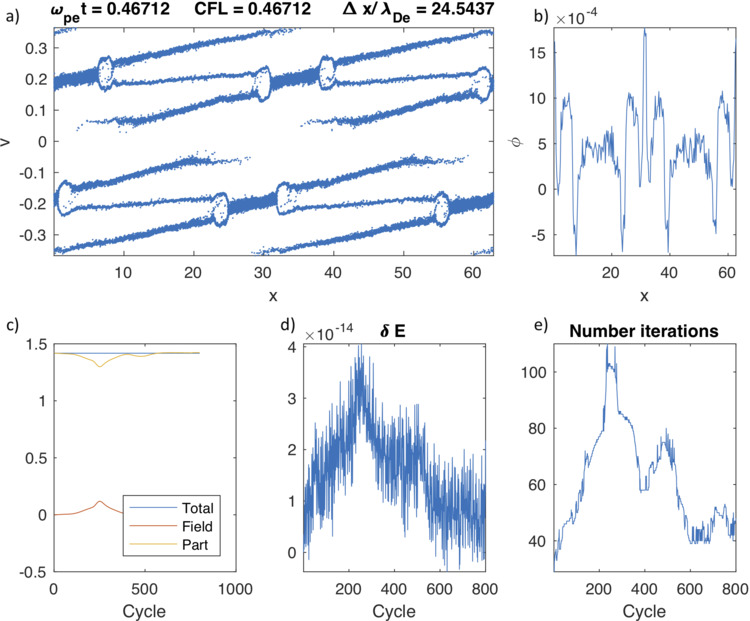}
	\caption{Method 2: Summary of a benchmark calculation, the final state is shown. Panel a: phase space (particle positions and velocities). Panel b: the potential $\varphi$. Panel c: Evolution of the energy. Panel d: Error in the conservation of energy. Panel e: Number of iterations needed for the NK iteration.}
	\label{fig:lin-vanillajfnk}
\end{figure}

The same case is repeated with Method 2 where the non-linearity of the field equations is simplified by using the potential at the $n$ time level. Figure \ref{fig:lin-vanillajfnk} shows the results. As can be seen, the differences are in panel a where the distribution function is shown are visible only in small details. Similar conclusions can be reached for panels b where the potential is also similar, with the same peaks (although the structure of the peaks varies slightly). There are clearly the same shock and same patterns. The differences however are quite significant in the number of iterations needed: the more non-linear case requires obviously more iterations.

But is the solution correct? We have two indications that it is. First, we have conducted a convergence study varying the number of cells and particles. The results presented are converged in the sense that the location of the features in phase-space does not change. Second, the presence of shock-like features in conditions similar to those reported here also led to similar phase space features~\citep{forslund1970formation}. We reached therefore confidence that teh solution is correct.

\begin{figure}[h]
	\centering
	\begin{tabular}{cc}
		a) Method 1  & b) Method 2 \\
\includegraphics[width=0.45\textwidth]{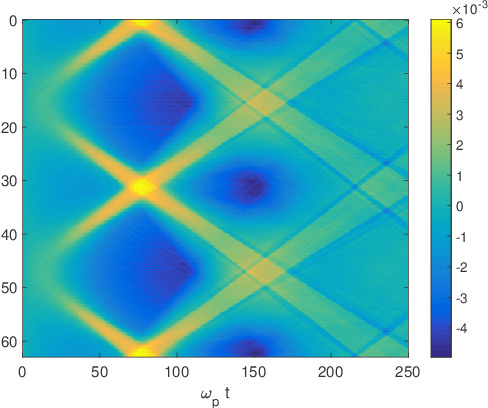}
&\includegraphics[width=0.45\textwidth]{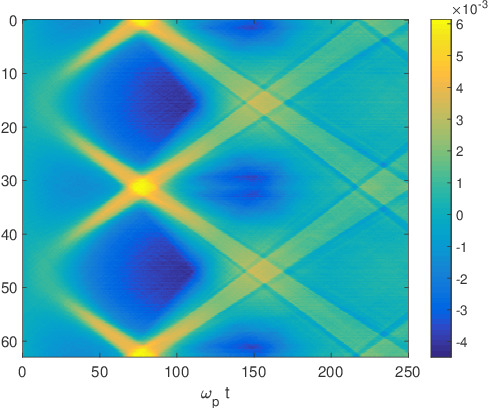}
	\end{tabular}
	\caption{False colour plot of the space-time evolution of the potential $\varphi$. Panel a (left) shows the results of Method 1 and panel b (right) of Method 2. Shocks are identified by the sharp transition and their reflection and interaction by the peak in the potential (bright yellow spots).}
	\label{fig:shock}
\end{figure}

These shocks travel through the system.  Figure  \ref{fig:shock} reports for both methods, the space-time evolution:  shocks can be identified by the sharp transition in the field value. The shocks move at constant speed. The periodicity of the boundary conditions allows a shock to exit one side an re-enter the other. Shocks are also observed to interact and pass through each other.

Besides the interesting physics accurately resolved, the point of the test is evaluating energy conservation. Energy is being exchanged between particles, medium and fields but the total energy is exactly conserved, see Fig. \ref{fig:vanillajfnk}-c and \ref{fig:lin-vanillajfnk}-c. The solution requires a tolerance in the iteration of the NK method. We set a tolerance of $10^{-14}$ and we indeed find energy conservation to be within the tolerance set, as shown in Fig. \ref{fig:vanillajfnk}-d and Fig. \ref{fig:lin-vanillajfnk}-d. If the tolerance level is modified so is the level of confidence on the energy conservation.  Both methods considered produce exact energy conservation.

For the two cases reported, the number of iterations at each cycle is shown in Fig. \ref{fig:vanillajfnk}-e and Fig. \ref{fig:lin-vanillajfnk}-e. The number of iterations has a remarkable increase corresponding to the time when the shocks are first formed. Method 1 with the explicit  nonlinearity of the field equation requires more iterations by a factor of approximately 50\% more, a non-negligible effect.

In the comparison above of Method 1 and Method 2 we have used the JFNK non-linear solver. The physical results are independent of the non-linear solver but the computational cost of reaching those results depend on the non-linear solver used: finding the most efficient is the target of the next section.

\section{Performance of Field-Hiding and Direct Newton Schwarz} 
As described above, the performance of the JFNK might be increased  if field-hiding or the DNS method is used. We focus now on Method 2 for two reasons. First, it was shown to be accurate and more computationally efficient in the test above. Second, it is simpler to implement field-hiding when the equations for the fields are formally linear. 

The latter point requires clarification. When field-hiding is used, for each Newton iteration applied to the particles, the fields need to be computed using the current Newton iterate of the particle velocity (and indirectly position). This operation can be done with a linear solver if the field equations are linear but it can also be done with another non-linear solver if the equations are non-linear. The first method nests a linear solver inside the function evaluation for the residual of the particles, the second nests a non-linear (for example another independent JFNK) solver for the field equation. Both can be done but of course the first is simpler.

To evaluate the performance of the three non-linear solvers, we consider a series of progressively larger systems. The following parameters are held fixed in all runs: time step $\Delta t=1.25$, the grid spacing  $\Delta x=20 \pi/64$ using $10000/64$ particles per cell, medium temperature  $e/kT =253.3030$. We change however the system size by multiples $M$ of $L=20\pi M$. We then use $10000 M$ particles and $64 M$ cells in each run. As $M$ is varied, we compare the number of iterations and the total CPU time for the direct solution with JFNK including both field  and particle residual (referred to as vanilla JFNK) with that of field-hiding and DNS. 

\begin{figure}
	\centering
	\includegraphics[width=0.7\linewidth]{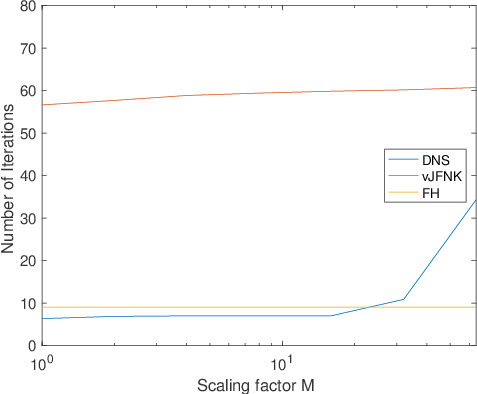}
	\caption{Scaling study for the number of Newton iterations needed for the vanilla JFNK and the field-hiding strategy. }
	\label{fig:scalingiter}
\end{figure}

Figure \ref{fig:scalingiter} shows the average number of Newton iterations needed for the three solution strategies. The two methods based on the JFNK iteration perform nearly ideally, with no significant variation on the number of Newton iterations needed as $M$ increases. The DNS performs well only for small systems, when the scaling $M$ increases the number of iterations balloons and the method fails. This is due to the slow convergence caused by the approximation of the Jacobian. 

The difference in iterations between FH and vJFNK is strong and compelling. The FH solution is clearly preferable. 

The number of iterations correspond to a similar nearly ideal performance in CPU time for vJFNK and FH. Again the DNS looses ground on larger systems.  FH results in almost an order of magnitude gain in computational time.

\begin{figure}
	\centering
	\includegraphics[width=0.7\linewidth]{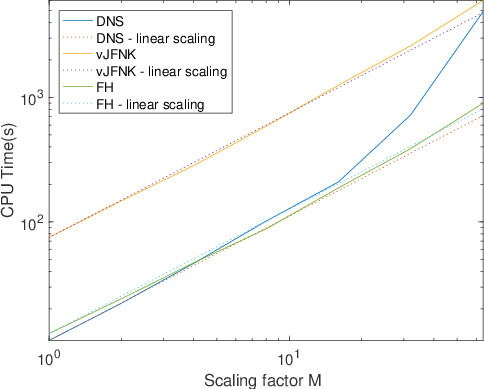}
	\caption{Scaling study for the CPU time needed for the vanilla JFNK and the field-hiding strategy. The ideal linear scaling is reported as dotted lines.}
	\label{fig:scalingcpu}
\end{figure}

All tests were done on a dual Intel(R) Xeon(R) CPU E5-2630 0 at 2.30GHz with cache size 15360 KB with 64GB RAM memory.  Note that in the analysis above we used the same tolerance for all three non-linear solvers to enforce the exact same accuracy in all three solutions that are therefore indistinguishable from the physical point of view. The difference is only in the number of iterations and the cost of the simulation. But the accuracy of the end result is the same. This is especially true for the DNS where the Jacobian is approximated. Approximating the Jacobian in the Newton method can slower the rate of convergence and the direction used to compute the next guess might become sub-optimal. However, the solution obtained is still exact, it is just might require more iterations.

\section{Conclusions}
We extended the recent developments on fully implicit energy conserving particle methods to the case where long-range forces are mediated by a medium with very different time and space scale than the inter-particle forces. This is the case of the Poisson-Boltzmann equation that describes a variety of physical systems.

We designed a new implicit method that is shown to conserve energy exactly to round off. The method is based on the solution with the JFNK method of non-linear system of equations composed by the momentum equation for each particle (Newton's law) and the equation of the medium (Boltzmann factor included in Poisson's equation). The challenge of the new approach is the computational cost. We investigate two strategies for reducing the number of iterations needed by hiding the field solution within the residual evaluation of the particles. 

In the FH approach, we still use the JFNK methods but applied only to the particle residual. At each new Newton iteration for the particle velocities provided the fields are computed directly (using a direct solver). This avoids the need to include also the field equations as part of the residual evaluation (a strategy referred to as field hiding).

In the DNS approach, even the JFNK for the particles is removed. We compute analytically the Schwarz decomposed Jacobian of each particles and invert it analytically. The coupling between particles is handled as a Picard iteration. The method does not require any extra storage besides one copy of the particle information. This is a strong reduction from the JFNK approach that needs to keep in memory multiple Krylov vectors with the same dimension of the residual (that is the size of the number of particle unknowns).

The three methods differ only in the solution strategy and the number of iterations but the solution is the same. 

We tested our approach  in  the case of two ion beams moving in a medium of Boltzmann-distributed electrons: a configuration leading to multiple interacting double layers (electrostatic shocks). A very taxing test for the methods. The field equations have been discretized in two ways, one with a stronger non-linearity that assumes the medium response at mid time level (Method 1) and one with weaker non-linearity that treats the medium at the beginning of the time level (Method 2). 

We show both to conserve energy and to lead to virtually identical results, but with a distinct advantage for the weaker non-linear discretization that evaluated the medium at the beginning of the time step.

We then focused on the Method 2 and compared the JFNK approach based on including all residuals in the evaluation with FH and DNS. Both methods based on a JFNK approach show a nearly ideal scaling, with the number of iterations remaining independent of the system size and the computational cost increasing linearly. On the other hand, the DNS starts to fail as the system size increases:  the number of iterations increase and the CPU time increases superlinearly.

\bibliographystyle{apalike}
\bibliography{lapenta}

\begin{thebibliography}{}

\bibitem[Birdsall and Langdon, 2004]{birdsall-langdon}
Birdsall, C. and Langdon, A. (2004).
\newblock {\em Plasma Physics Via Computer Simulation}.
\newblock Taylor \& Francis, London.

\bibitem[{Brackbill} and {Cohen}, 1985]{Brackbill:1985}
{Brackbill}, J.~U. and {Cohen}, B.~I., editors (1985).
\newblock {\em {Multiple time scales.}}

\bibitem[Brackbill and Cohen, 2014]{brackbill2014multiple}
Brackbill, J.~U. and Cohen, B.~I. (2014).
\newblock {\em Multiple time scales}, volume~3.
\newblock Academic Press.

\bibitem[Chen et~al., 2011]{chen2011energy}
Chen, G., Chac{\'o}n, L., and Barnes, D.~C. (2011).
\newblock An energy-and charge-conserving, implicit, electrostatic
  particle-in-cell algorithm.
\newblock {\em J. Comput. Phys.}, 230(18):7018--7036.

\bibitem[Chen et~al., 2012]{chen2012efficient}
Chen, G., Chac{\'o}n, L., and Barnes, D.~C. (2012).
\newblock An efficient mixed-precision, hybrid cpu--gpu implementation of a
  nonlinearly implicit one-dimensional particle-in-cell algorithm.
\newblock {\em J. Comput. Phys.}, 231(16):5374--5388.

\bibitem[Chen et~al., 2014]{chen2014fluid}
Chen, G., Chac{\'o}n, L., Leibs, C.~A., Knoll, D.~A., and Taitano, W. (2014).
\newblock Fluid preconditioning for newton--krylov-based, fully implicit,
  electrostatic particle-in-cell simulations.
\newblock {\em J. Comput. Phys.}, 258:555--567.

\bibitem[De~Boor et~al., 1978]{de1978practical}
De~Boor, C., De~Boor, C., Math{\'e}maticien, E.-U., De~Boor, C., and De~Boor,
  C. (1978).
\newblock {\em A practical guide to splines}, volume~27.
\newblock Springer-Verlag New York.

\bibitem[de~Borst, 2008]{de2008challenges}
de~Borst, R. (2008).
\newblock Challenges in computational materials science: Multiple scales,
  multi-physics and evolving discontinuities.
\newblock {\em Computational Materials Science}, 43(1):1--15.

\bibitem[Eastwood et~al., 1984]{eastwood1984p3m3dp}
Eastwood, J., Hockney, R., and Lawrence, D. (1984).
\newblock P3m3dp-the three-dimensional periodic particle-particle/particle-mesh
  program.
\newblock {\em Computer Physics Communications}, 35.

\bibitem[Essmann et~al., 1995]{essmann1995smooth}
Essmann, U., Perera, L., Berkowitz, M.~L., Darden, T., Lee, H., and Pedersen,
  L.~G. (1995).
\newblock A smooth particle mesh ewald method.
\newblock {\em The Journal of chemical physics}, 103(19):8577--8593.

\bibitem[Fogolari et~al., 2002]{fogolari2002poisson}
Fogolari, F., Brigo, A., and Molinari, H. (2002).
\newblock The poisson--boltzmann equation for biomolecular electrostatics: a
  tool for structural biology.
\newblock {\em Journal of Molecular Recognition}, 15(6):377--392.

\bibitem[Forslund and Shonk, 1970]{forslund1970formation}
Forslund, D. and Shonk, C. (1970).
\newblock Formation and structure of electrostatic collisionless shocks.
\newblock {\em Physical Review Letters}, 25(25):1699.

\bibitem[Frenkel and Smit, 2001]{frenkel2001understanding}
Frenkel, D. and Smit, B. (2001).
\newblock {\em Understanding molecular simulation: from algorithms to
  applications}, volume~1.
\newblock Elsevier.

\bibitem[Grigoryev et~al., 2002]{GrigoryevPIC}
Grigoryev, Y.~N., Vshivkov, V.~A., and Fedoruk, M.~P. (2002).
\newblock {\em Numerical Particle-in-Cell Methods: Theory and Applications}.
\newblock de Gruyter.

\bibitem[Hockney and Eastwood, 1988]{hockney-eastwood}
Hockney, R. and Eastwood, J. (1988).
\newblock {\em Computer simulation using particles}.
\newblock Taylor \& Francis.

\bibitem[Honig and Nicholls, 1995]{honig1995classical}
Honig, B. and Nicholls, A. (1995).
\newblock Classical electrostatics in biology and chemistry.
\newblock {\em Science}, 268(5214):1144--1149.

\bibitem[Im et~al., 1998]{im1998continuum}
Im, W., Beglov, D., and Roux, B. (1998).
\newblock Continuum solvation model: computation of electrostatic forces from
  numerical solutions to the poisson-boltzmann equation.
\newblock {\em Computer physics communications}, 111(1-3):59--75.

\bibitem[Kelley, 1995]{kelley}
Kelley, C.~T. (1995).
\newblock {\em Iterative methods for linear and nonlinear equations}.
\newblock SIAM, Philadelphia.

\bibitem[Kim et~al., 2005]{kim}
Kim, H., Chac{\`o}n, L., and Lapenta, G. (2005).
\newblock Fully implicit particle-in-cell algorithm.
\newblock {\em Bull. Am. Phys. Soc.}, 50:2913.

\bibitem[Knoll and Keyes, 2004]{knoll2004jacobian}
Knoll, D.~A. and Keyes, D.~E. (2004).
\newblock Jacobian-free newton--krylov methods: a survey of approaches and
  applications.
\newblock {\em J. Comput. Phys.}, 193(2):357--397.

\bibitem[Koehl, 2006]{koehl2006electrostatics}
Koehl, P. (2006).
\newblock Electrostatics calculations: latest methodological advances.
\newblock {\em Current opinion in structural biology}, 16(2):142--151.

\bibitem[Lapenta and Markidis, 2011]{lapenta2011particle}
Lapenta, G. and Markidis, S. (2011).
\newblock Particle acceleration and energy conservation in particle in cell
  simulations.
\newblock {\em Physics of Plasmas (1994-present)}, 18(7):072101.

\bibitem[Lu et~al., 2008]{lu2008recent}
Lu, B., Zhou, Y., Holst, M., and McCammon, J. (2008).
\newblock Recent progress in numerical methods for the poisson-boltzmann
  equation in biophysical applications.
\newblock {\em Commun Comput Phys}, 3(5):973--1009.

\bibitem[Markidis and Lapenta, 2011]{markidis2011energy}
Markidis, S. and Lapenta, G. (2011).
\newblock The energy conserving particle-in-cell method.
\newblock {\em J. Comput. Phys.}, 230(18):7037--7052.

\bibitem[Morrone et~al., 2010]{morrone2010molecular}
Morrone, J.~A., Zhou, R., and Berne, B. (2010).
\newblock Molecular dynamics with multiple time scales: how to avoid pitfalls.
\newblock {\em Journal of chemical theory and computation}, 6(6):1798--1804.

\bibitem[Salmon and Warren, 1994]{salmon1994skeletons}
Salmon, J.~K. and Warren, M.~S. (1994).
\newblock Skeletons from the treecode closet.
\newblock {\em Journal of Computational Physics}, 111(1):136--155.

\bibitem[Siddi et~al., 2018]{siddicomparison}
Siddi, L., Cazzola, E., and Lapenta, G. (2018).
\newblock Comparison of preconditioning strategies in energy conserving
  implicit particle in cell methods.
\newblock {\em Communication in Computational Physics}, 24(3):672--694.

\bibitem[Tuckerman et~al., 1992]{tuckerman1992reversible}
Tuckerman, M., Berne, B.~J., and Martyna, G.~J. (1992).
\newblock Reversible multiple time scale molecular dynamics.
\newblock {\em The Journal of chemical physics}, 97(3):1990--2001.

\end{thebibliography}
\end{document}